\documentclass[%
 reprint,
 amsmath,amssymb,
 aps,
]{revtex4-1}

\usepackage{graphicx}
\usepackage{dcolumn}
\usepackage{bm}

\begin{document}

\preprint{APS/123-QED}

\title{Mechanical Unfolding of a Simple Model Protein Goes Beyond the Reach of One-Dimensional Descriptions}

\author{R. Tapia-Rojo}%
\email{Rafa.T.Rojo@gmail.com}
\affiliation{Instituto de Biocomputaci\'on y F\'{\i}sica de Sistemas Complejos and Departamento de F\'{i}sica de la Materia Condensada,  Universidad de Zaragoza, 50009 Zaragoza, Spain}

\author{S. Arregui}%
\affiliation{Instituto de Biocomputaci\'on y F\'{\i}sica de Sistemas Complejos and Departamento de F\'{i}sica de la Materia Condensada,  Universidad de Zaragoza, 50009 Zaragoza, Spain}

\author{J.J. Mazo}%
\affiliation{Instituto de Ciencia de Materiales de Arag\'on and Departamento de F\'{i}sica de la Materia Condensada, CSIC-Universidad de Zaragoza, 50009 Zaragoza, Spain}

\author{F. Falo}%
\affiliation{Instituto de Biocomputaci\'on y F\'{\i}sica de Sistemas Complejos and Departamento de F\'{i}sica de la Materia Condensada,  Universidad de Zaragoza, 50009 Zaragoza, Spain}

\date{\today}

\begin{abstract}
We study the mechanical unfolding of a simple model protein. The
Langevin dynamics results are analyzed using Markov-model methods
which allow to describe completely the configurational space of the
system. Using transition path theory we also provide a quantitative description of
the unfolding pathways followed by the system. Our study shows a complex dynamical scenario. In particular,
we see that the usual one-dimensional picture: free-energy vs
end-to-end distance representation, gives a misleading description of
the process. Unfolding can occur following different pathways and
configurations which seem to play a central role in one-dimensional
pictures are not the intermediate states of the unfolding dynamics.

\begin{description}
\item[PACS numbers]
87.15.hm, 87.15.A-, 87.15.H-, 87.15.La
\end{description}
\end{abstract}

\pacs{Valid PACS appear here}
\maketitle


\section{Introduction}

The characterization of folding and unfolding energy landscapes of
biomolecules is a major problem in
biophysics which sheds light onto
biomolecules' role and function~\cite{unf0,unf1,unf2,unf3,unf5}.  In
this effort, the emergence of single-molecule techniques that let the
manipulation of individual molecules has opened a new wide field,
allowing to monitor unfolding processes by looking into a single
specimen~\cite{AFM1,AFM2,AFM3,OT1,OT2,MT1,MT2,FRET1,FRET2}.

In force-pulling experiments, the one-dimensional description is
usually adopted, as force is considered to impose a preferred
direction that appears as the slowest degree of freedom compared with
the remaining ones. In this sense, optical tweezers~\cite{OT1,OT2},
magnetic tweezers~\cite{MT1,MT2} or AFM~\cite{AFM1,AFM2,AFM3}
experiments are usually analyzed considering the end-to-end distance
as the proper reaction coordinate, with a well developed force
spectroscopy theory~\cite{dudko06,Dudko2008, dudko11, review2014} that allows stating
predictions grounded on this hypothesis. Also, recent studies of
single molecule Foester resonant energy transfer fluorescence
study thermal unfolding by tracking the radius of gyration of
individual molecules~\cite{FRET1,FRET2}.  Computational works
similarly take advantage of this simple description, choosing reaction
coordinates such as the fraction of native contacts
$Q$~\cite{Q1,Q2,Best13,Best132}, the RMSD from the native
structure~\cite{RMSD1} or the Principal Components~\cite{stock07, PCA93, PCA92, pca2d}.
Nevertheless, this tempting approach must be used with great care, as
some energy minima which represent relevant metastable conformations
and the barriers connecting such states may be hidden when projecting
the actual large-dimensional free energy landscape onto a
low-dimensional subspace. Besides, one dimensional profiles might
suggest misleading unfolding paths, consequence of this projection
restriction.

In order to explore such aspects, we choose a coarse-grained model
protein~\cite{BLN1, BLN2, modelor,modelor2, wales3, wales4,
  model,model2,model3, wales2} and study it through a force-clamp
protocol. The output of the simulations will be analyzed through two
different approaches, allowing a comparison between the conclusions
yielded by each. First, we build one dimensional free-energy profiles
along the end-to-end distance and the fraction of native
contacts. Second, we describe the configurational space of the system
by using Markov-Model methods \cite{noebook, pande1, cmn1, dpg} and
obtain the unfolding paths applying transition-path theory \cite{TPT1,
  TPT2, TPT3, TPT4}.

Although recent works cast doubt on a simple low dimensional
description of thermal (un)folding processes~\cite{stock07, multi}, the
one-dimensional approach is usually adopted for mechanical unfolding
processes, due to the privileged direction imposed by the force
\cite{multi2, dudko06}. In the case studied here, this fact, together
with the simplicity of the protein structure, apparently point to a
valid one-dimensional description of the unfolding
process. Nevertheless, we find out that one-dimensional profiles lead
to deceptive conclusions.  In particular, these profiles suggest the
existence of a metastable state (the half-stretched configuration, see
Fig.~\ref{PMF}) as a mechanical intermediate between the native and
stretched states. Opposed to this, we find that unfolding occurs
through two major routes defined by the existence of two different
mechanical intermediates, not identified in the one-dimensional
description. Although very stable, the half-stretched configuration
plays a marginal role in the unfolding process. This multi-path
picture can never be captured through a one-dimensional
description. In addition we are able to systematically define all the
individual unfolding pathways calculating their relative weight in the
dynamics and yielding a complete and quantitative vision of the
protein's landscape that completes the picture described in previous
studies on the same system~\cite{model,model2,model3}.

\section{Model}

The BLN model \cite{BLN1, BLN2} is a coarse grained off-lattice protein model
in which the residues are represented by ``colored" beads, hydrophobic (B), hydrophilic (L)
and neutral (N). Due to its rich behavior, despite its simplicity, this model has been widely studied, with several modifications through time \cite{modelor, modelor2, wales3, model, model2, model3}. In particular, the $46$-residue sequence 
(BLN-$46$) $B_9N_3(LB)_4N_3B_9N_3(LB)_5L$ folds into a four-strand $\beta$ barrel
showing nonetheless a frustrated ground state \cite{wales3}.

The potential terms we use account for a stiff nearest-neighbor harmonic potential, 
a three-body bending interaction, a four-body dihedral interaction and a sequence dependent
Lennard-Jones potential \cite{model2, model3}:
\begin{eqnarray}
V_{BLN}&=&\frac{1}{2}K\sum_{i=1}^{N-1}(r_{i,i+1}-r_0)^2\\ \nonumber
&+&\sum_{i=1}^{N-2}\left[A\cos\theta_i+B\cos2\theta_i-V_0\right]\\ \nonumber
&+&\sum_{i=1}^{N-3}\left[C_i(1+\cos\phi_i)+D_i(1+\cos3\phi_i)\right]\\ \nonumber
&+&\sum_{ij}\epsilon_{ij}\left(\frac{1}{r^{12}_{ij}}-\frac{c_{ij}}{r^6_{ij}}\right),
\end{eqnarray}
\noindent where $r_{ij}$ is the distance between residues $i$ and $j$, $\theta$ is the bending angle and
$\phi$ the dihedral angle. For parameter values see \cite{model2} and Appendix A.

We simulate the system by integrating Langevin equations of motion at
constant temperature $T$ and following a force-clamp protocol, where monomer $1$ is fixed 
while a constant force is applied to the last monomer, $46$, through a linear spring. Such equations are given by
\begin{equation}
m\mathbf{\ddot{r}}_i=-\gamma \mathbf{\dot{r}}_i-\nabla_i V_{BLN}+\mathbf{F}_i+\boldsymbol{\eta}_i,
\end{equation}
\noindent where $m$ is each residue unitary mass, $\gamma$ the friction coefficient, $\mathbf{F}$ the external force applied in the $z$ direction and $\eta_i$ 
Gaussian white noise of zero average, holding fluctuation-dissipation theorem $\langle \eta_i\eta_j\rangle=2T\gamma\delta(t-t')\delta_{ij}$.

This model protein has a well characterized unfolding transition (see \cite{model2} and Appendix C) at $T_c$
and unfolds mechanically at $F_U$. We work from now on at $T=0.55T_c$ and $F=0.8F_U$ in order to maximize the number of 
configurations visited by the system. Lower forces would not populate the unfolded state while above $F_U$ the unfolding would be irreversible.

\section{Methods}
We present here the different methods use to analyze the simulated
trajectories in order to understand the mechanical unfolding scenario
of our model system.

\subsection{Potential of Mean Force}
The Potential of Mean Force (PMF) is a low dimensional (typically one-dimensional) characterization
of the free energy landscape of a system, which relies on the choice of a reaction 
coordinate $X$. The PMF is simply 
$F/k_BT=-\log P(X)$, where $P(X)$ is the probability density of the chosen
reaction coordinate $X$.

We will explore the PMF  of the system (section IV.A) by using two different reaction coordinates.
As the mechanical force imposes a privileged direction,  the end-to-end distance 
$\xi=|\mathbf{r}_N-\mathbf{r}_0|$ appears as a natural choice. This magnitude is indeed
widely used in most single molecule force spectroscopy applications \cite{dudko06, app1, app2, app3}. Additionally,
we use the fraction of native contacts $Q$ \cite{Q1,Q2}, often reported in computational applications
as a good magnitude for describing protein unfolding, based on the importance of topology
on protein structure. 

\subsection{Principal Components Analysis}
Principal Component Analysis (PCA) is a standard statistical  method for reducing
the dimensionality of a complex system such as biological molecule \cite{PCA92, PCA93, pca2d}. PCA performs a linear transformation by diagonalizing
the covariance matrix $\mathcal{C}_{ij}=\langle y_i y_j\rangle -\langle y_i\rangle\langle y_j\rangle$,
removing thus all internal correlations. The Principal Components (PCs) $q_i$ are calculated as the projection of the trajectory onto each eigenspace.
If we order the eigenvalues, the first largest PCs
contain most of the fluctuations of the system and can be used as adequate reaction coordinates.

\subsection{Conformational Markov Network}
In order to characterize the thermodynamical and kinetic properties of our system we
build a Markov Model~\cite{noebook,pande1} by discretizing the state space of our molecule into a set
$S=\{1, 2, \cdots, M\}$ of $M$ conformational states  defining the Conformational Markov Network of the system \cite{cmn1, dpg}.
 For our system, the conformational space is defined as the first three PCs,
reducing greatly its dimensionality  but keeping its essential features.
With these three coordinates we maintain the $75\%$ of the system fluctuations, while the remaining ones
account for symmetric thermal fluctuations. Each of the coordinates is discretized into $30$ bins of equal
volume, thus $M=27000$.

The Conformational Markov Network is built from the dynamical trajectories, by counting the occupation
of each of the states $\pi_i$ and calculating the transition matrix $T_{ij}$ which measures the probability of
going from state $i$ to state $j$ within time $\tau$, being $\tau$ the time window or lag time used to analyze our
trajectories ($\tau=15ps$ in our case).

The transition matrix $\tilde{T}$ is ergodic and, if the molecule is in equilibrium, the occupation distribution $\pi_i$
can be recovered as the eigenvector with eigenvalue $1$. In such situation, detail balance condition holds, $\pi_iT_{ij}=\pi_jT_{ji}$,
and $\pi$ is the Boltzmann distribution.

\subsection{Basins of attraction Network}
As the Conformational Markov Network is typically made up of thousands of nodes and links,
hardly any relevant physical information can be directly obtained. A clustering or coarse-graining process
is usually followed in order to group together nodes with similar physical features leading to an smaller, 
more meaningful network.

Here we apply the Stochastic Steepest Descent algorithm \cite{dpg} (see Appendix B.2 for detailed algorithm). 
The advantage of this algorithm is that the network is systematically split into its basins of attraction \emph{i.e.} 
groups of nodes whose probability flux converges into a single node (minimum). The coarse-graining process does 
not rely in any arbitrary definition, but on the kinetic properties of the system.
Physically, while each node would represent microstates of the system, the basins of attraction represent macrostates.

Onto this network we calculate a new transition matrix ${T}_{ij}$ and the occupation probability of each basins $\pi_i$.
Free energy differences from basin $i$ and $j$ are given by $\Delta F_{ij}=-k_BT\log\pi_i/\pi_j$. The mean escape time
from basin $i$ is defined as $\langle t_s\rangle =\tau/(1-T_{ii})$, where $\tau$ is the time window used to sample the configurations,
while transition times between basins $i$ and $j$ are defined as $\tau_{i\rightarrow j}=\tau/T_{ij}$.

\subsection{Transition-Path Theory}

The Markov Network defined above contains all  thermodynamic and kinetic information of the system. Nevertheless,
we are interested in computing the transition pathways between the set of native conformations to the
fully stretched conformation. Transition-Path theory provides the necessary tools for doing this \cite{TPT1, TPT2, TPT3}.
We define $A$ as the subset of basins which represent the native conformation while $B$ is the subset of stretched basins. 
Our question is which is the typical sequence of intermediate $I$ states to go from $A$ to $B$.

The committor probability $q^+_i$ is defined as the probability, when starting at state $i$, to reach set $B$ next 
rather than $A$. In our case, this is the unfolding probability. By definition $q^+_i=0$ if $i\in A$ and $q^+_i=1$ if
$i\in B$. Mathematically, the committor probability can be computed by solving the following system of linear equations:
\begin{equation}
-q^+_i+\sum_{k\in I}T_{ik}q^+_k=-\sum_{k\in B}T_{ik}.
\end{equation}

For a molecule in equilibrium, the backward-committor probability $q^-_i$ is simply $q^-_i=1-q^+_i$.

The transition matrix $T_{ij}$ contains information from every possible trajectory which appears in the 
equilibrium ensemble of the molecule. In order to extract the contributions from the unfolding trajectories $A\rightarrow B$, 
we calculate the effective flux $f_{ij}$ defined as the probability flux from $i\rightarrow j$ contributing to the $A\rightarrow B$ transition:
\begin{equation}
f_{ij}=\pi_iq^-_iT_{ij}q^+_j.
\end{equation}

If we want to calculate the unfolding flux, removing recrossings which might appear in a $A\rightarrow B$ transition, we need to define  the net flux as
\begin{equation}
f^+_{ij}=\max[0, f_{ij}-f_{ji}],
\end{equation}

\noindent $f^+_{ij}$ defines a network of fluxes that go from $A$ to $B$. The total unfolding flux $F$ represents the expected number of $A\rightarrow B$ transitions per time window
$\tau$ and is defined as:
\begin{equation}
F=\sum_{i\in A}\sum_{j\notin A}\pi_iT_{ij}q^+_{j}.
\end{equation}

 In order to decompose this flux network onto individual pathways $P_i$, 
different approaches can be applied \cite{TPT3, TPT4}. Here we base our strategy on the bottleneck algorithm, where given an individual pathway, 
the bottleneck (rate limiting step) is identified as the minimal net flux of the path $f_i$ and subtracted from every remaining net flux $f^+_{ij}$.
The process is iterated until the network is fully decomposed into a set of individual pathways $P_i$.

\section{Results}
In order to elucidate the unfolding mechanism under the effect of mechanical force for our model protein,
we have performed six long equilibrium simulations. Every simulation starts from the native configuration,
is equilibrated for $3\mu s$ and then runs up to $3 ms$.

\subsection{One dimensional description: the Potential of Mean Force}

\begin{figure}[h!]
\begin{center}
\includegraphics[width=0.5\textwidth]{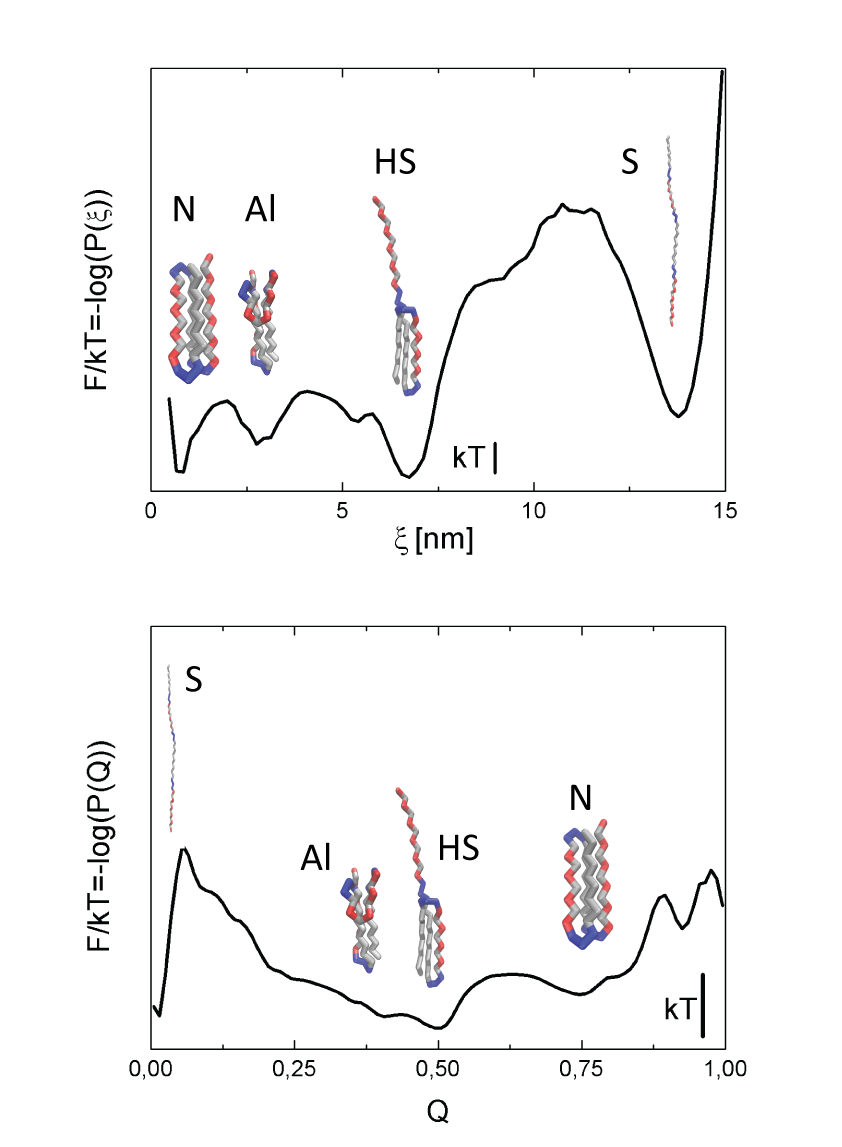}
\caption{\footnotesize Potential of Mean Force as a function of the
  end-to-end distance $\xi$ and the fraction of native contacts $Q$.}
\label{PMF}
\end{center}
\vspace{-0.7cm}
\end{figure}

Figure~\ref{PMF} shows the PMF calculated along the end-to-end
distance $\xi$ and the fraction of native contacts $Q$ of our model
protein. The profile for $\xi$ shows four clear minima that can be
identified with four different configurations, considering that each of the
$\beta$ strands has a length of $\xi\sim 3\,nm$. In the native configuration
($N$) $\xi\sim0\,nm$, as the extremal $\beta$ strands are oriented
in the same direction. In the aligned configuration ($Al$) the second
strand $(LB)_4$ is bent so that the extremes are aligned in the
pulling direction and $\xi\sim 3\,nm$. The half-stretched
configuration ($HS$) appears as an stable minima at $\xi\sim6\,nm$,
as the fourth $(LB)_5$ strand is unfolded. The fully stretched
configuration ($S$), with $\xi\sim 12\,nm$, shows the protein
totally unfolded, as an stretched polymer.  

These states can also be identified in the $Q$ profile. State $S$ has all contacts broken
$Q\sim 0$, while  $Al$ and $HS$ maintain around half of the
contacts ($Q\sim0.5$). The $N$ configuration shows a minimum at $Q\sim 0.75$,
as thermal fluctuations break on average some of the contacts.
 
Remarkably, for this value of the force, the $HS$ configuration
correspond to the lowest minimum in both free energy profiles, and
thus is the most stable configuration. Its position in the PMF
suggests that it also has a relevant role in the stretching pathways,
appearing as a clear mechanical intermediate between the native and
fully stretched configuration.  In addition, it is necessary to jump
over a barrier of several $k_BT$ to reach state $S$ while the other
states are separated by low barrier. This suggest a fast dynamics
between $N$ and $HS$ and longer time scales to visit state $S$.

\subsection{Two dimensional description: Principal Component Analysis}

Before describing the Markov Model of the system, it is worth to exploit further the information PCA provides. As explained previously,
we build the Markov network by discretizing the first three PCs, which define our conformational space, with lower dimensionality, but still
capturing the main aspects of the system dynamics.

\begin{figure}[h!]
\begin{center}
\includegraphics[width=0.5\textwidth]{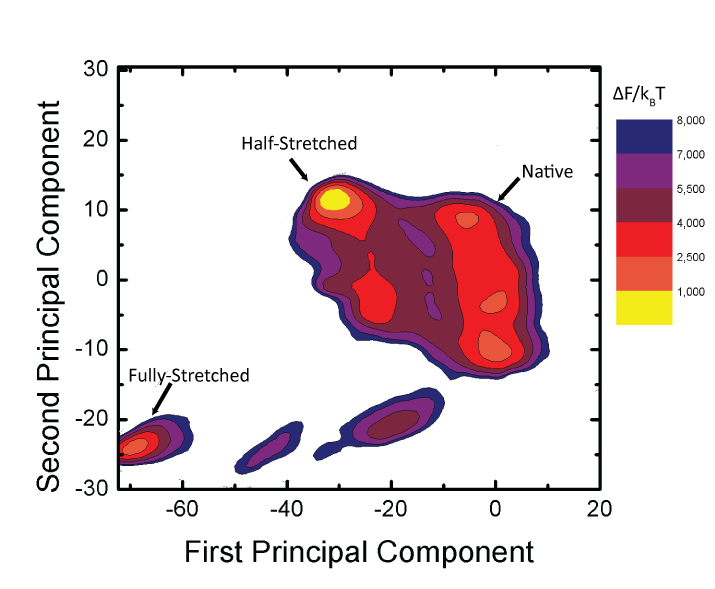}
\caption{\footnotesize Free-energy landscape along the first two PCs.}
\label{PCA}
\end{center}
\vspace{-0.7cm}
\end{figure}

Figure \ref{PCA} shows the free-energy landscape along the first two principal components $\Delta F/k_BT=-\log P(q_1,q_2)$. Its basic 
features agree with the one dimensional landscapes shown in previous section, as three major wells are found. Nevertheless we see also clear differences,
being the PCs able to capture better the details of the free-energy landscape. Each of these major wells have a rough structure, showing a set of minor wells
separated by small energy barriers $\sim 2k_BT$, revealing thus a richer variety of configurations. Moreover, two new low populated wells appear between the folded
structures (native and half-stretched) and the fully-stretched configurations. These new states could suggest the existence of different unfolding pathways,
where the half-stretched configuration does not necessarily plays the role of mechanical intermediate.

\subsection{Equilibrium ensemble of the model protein: the Basin Network }

The built microstate network is made up of $1876$ nodes related kinetically through $23995$ links.
After applying the Stochastic Steepest Descent algorithm \cite{dpg}, the network is clustered into $30$ basins
connected through $1290$ links. In order to obtain a good description of the system, we keep only those basins 
which were visited at least  $0.001\%$ of the trajectory ($\pi_i>10^{-5}$), avoiding pathological or extremely rare states.
After this refinements, we keep $13$ macrostates, connected through $65$ edges, including auto-links.

\begin{figure}[t!]
\begin{center}
\includegraphics[width=0.5\textwidth]{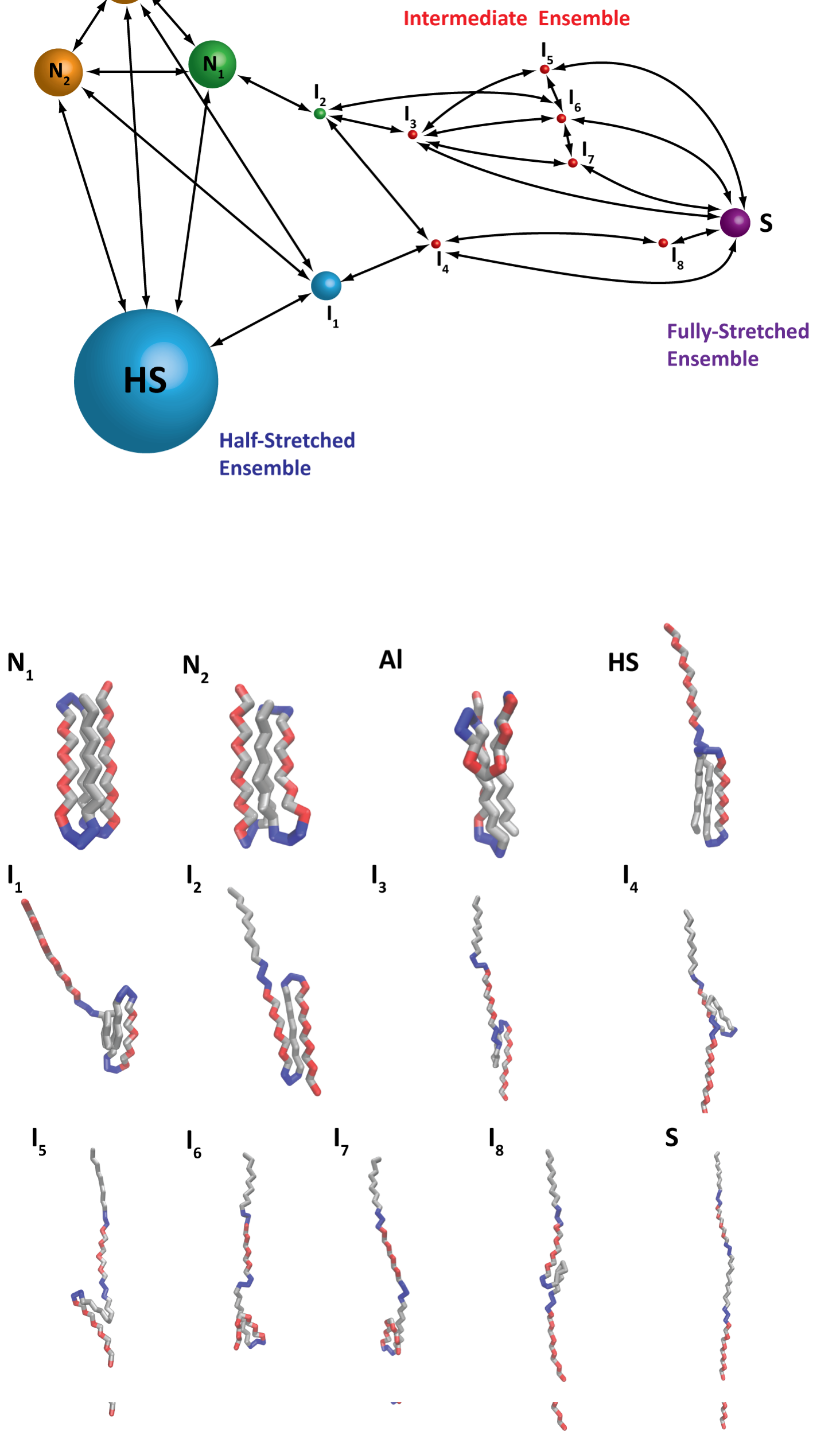}
\caption{\footnotesize Basins of attraction Markov Network (\emph{Upper}). We represent the $13$ basins with $\pi>10^{-5}$ where the size of the bead is proportional
to $\pi_i$. The bidirectional arrows connecting nodes represent allowed transitions (the magnitude of  $T_{ij}$ is not shown). Each basin is labelled according to the configuration
they encode. Representative structure associated to each basin (\emph{Lower}).}
\label{net}
\end{center}
\vspace{-0.7cm}
\end{figure}

Figure \ref{net} (upper) shows a graphical representation of the basin network, where the size of each bead (node)
is proportional to its occupation $\pi_i$. The spatial arrangement of the nodes was calculated applying the \emph{Force Atlas} algorithm \cite{gephi},
where an artificial dynamics is simulated. This dynamics is based in considering each link as a linear spring and including a certain repulsion between nodes, until an equilibrium 
configuration is obtained. The nodes are colored according to the modularity class they belong to \cite{modul}, having five different classes. 
Lower panel of Fig. \ref{net} shows a representative structure of each basin (macrostate), including the label which identifies
them.
 
Configurations $N_1$ and $N_2$ correspond to native-like states and
will define the native set $A$ due to its structural similarity and
high $Q$ value. The aligned configuration $Al$, already identified in
Fig.~\ref{PMF}, appears close to $N_1$ and $N_2$ in Fig.~\ref{net} but
does not belong to the native set since it gives very different $Q$
and $\xi$ values.  Basin $HS$ is the Half-Stretched Configuration, the
most stable macrostate under these conditions. State $S$ is the
Fully-Stretched Configuration, while the remaining $8$ basins are
labelled as intermediate states and will be discussed further on.

\begin{table}[h!]
\caption{\label{tabla}%
Description of the basins of attraction.
}
\begin{ruledtabular}
\begin{tabular}{cccccc}
\# & $\pi_i$ & $\langle t_s\rangle$ [ps] & $\langle Q\rangle$ & $\langle \xi \rangle$ [nm] & $q^+_i$\\
\hline
$N_1$ & $0.15$ & $559$ & $0.75$ & $0.8$ & $0.0$ \\
$N_2$ & $0.14$ & $495$ & $0.73$ & $0.9$ & $0.0$ \\
$Al$ & $0.14$ & $272$ & $0.40$ & $2.6$ & $1.4\times 10^{-4}$ \\
$HS$ & $0.44$ & $2982$ & $0.46$ & $6.5$ & $9.2\times 10^{-4}$ \\
$I_1$ & $0.07$ & $362$ & $0.25$ & $4.8$ & $1.2\times 10^{-3}$ \\
$I_2$ & $0.01$ & $2586$ & $0.35$ & $6.8$ & $0.12$ \\
$I_3$ & $6.67\times 10^{-5}$ & $120$ & $0.12$ & $9.0$ & $0.29$ \\
$I_4$ & $1.3\times 10^{-4}$ & $198$ & $0.11$ & $10.1$ & $0.34$ \\
$I_5$ & $1.9\times 10^{-5}$ & $64$ & $0.10$ & $9.6$ & $0.51$ \\
$I_6$ & $3.9\times 10^{-4}$ & $163$ & $0.14$ & $8.55$ & $0.53$ \\
$I_7$ & $3.3\times 10^{-4}$ & $176$ & $0.13$ & $9.35$ & $0.58$ \\
$I_8$ & $2.5\times 10^{-5}$ & $56$ & $0.09$ & $10.5$ & $0.71$ \\
$S$ & $0.06$ & $75000$ & $0.01$ & $13.7$ & $1$ \\
\end{tabular}
\end{ruledtabular}
\end{table}

Table \ref{tabla} shows information about each of the identified macrostates. $\pi_i$ is the occupation of basin $i$, 
$\langle t_s\rangle$ the mean escape time (defined above), $\langle Q \rangle$ the mean fraction of native contacts and $\langle \xi \rangle $ the mean
end-to-end distance, both calculated from the marginal distributions of such magnitudes on each basin. It is remarkable that in many cases
such distributions are not unimodal, so the actual meaning of the average must be taken with care. Finally, $q^+_i$ are the committor probabilities
from the native ($N_1$ and $N_2$) to the stretched ($S$) configuration this is: the unfolding probability of basin $i$.

It is important to stress the difference between the two native basins
$N_1$ and $N_2$, as they have very different connectivity features in
the network, belonging to different modularity classes. Configuration
$N_1$ is closer to the native structure, given the arrangement of the
neutral turns, while $N_2$ shows bigger fluctuations, leading to a
loss of some contacts. Interestingly, $N_1$ is more connected to the
Intermediate States than $N_2$, which shows fast transition times to
$HS$, $\tau_{N_2\rightarrow HS}=557ps$, while $\tau_{N_1\rightarrow
  HS}=13.5 \times 10^6ps$. In fact, they are both scarcely connected
-$\tau_{N_2\rightarrow N_1}=14\times 10^3 ps$ and
$\tau_{N_1\rightarrow N_2} =15\times 10^3 ps$-, reason why they belong
to a different modularity class. In this regard, in spite its
structural similarity which overlap both states in the PMF
description, their actual role in the configurational space is quite
different.

In this sense, the first contradictions with the conclusions yielded
by the PMF description appear here. While both descriptions agree
coarsely in the main features of the equilibrium ensemble of the
system, revealing three major states (native, half-stretched and
fully-stretched), the role of such states and the presence of other
relevant configurations is hidden in the one-dimensional
projection. $N_1$ and $N_2$ states are integrated into the same high
$Q$ or low $\xi$ minimum, will the intermediate low-populated states
which connect to the stretched state are impossible to be identified
in the one-dimensional representation.

\subsection{The unfolding pathways: Transition Path Theory}

In order to decipher the actual unfolding mechanism of our model protein under the effect of a mechanical force, we apply  Transition Path theory 
to the basin network, as explained in Methods section.

\begin{figure}[h!]
\begin{center}
\includegraphics[width=0.5\textwidth]{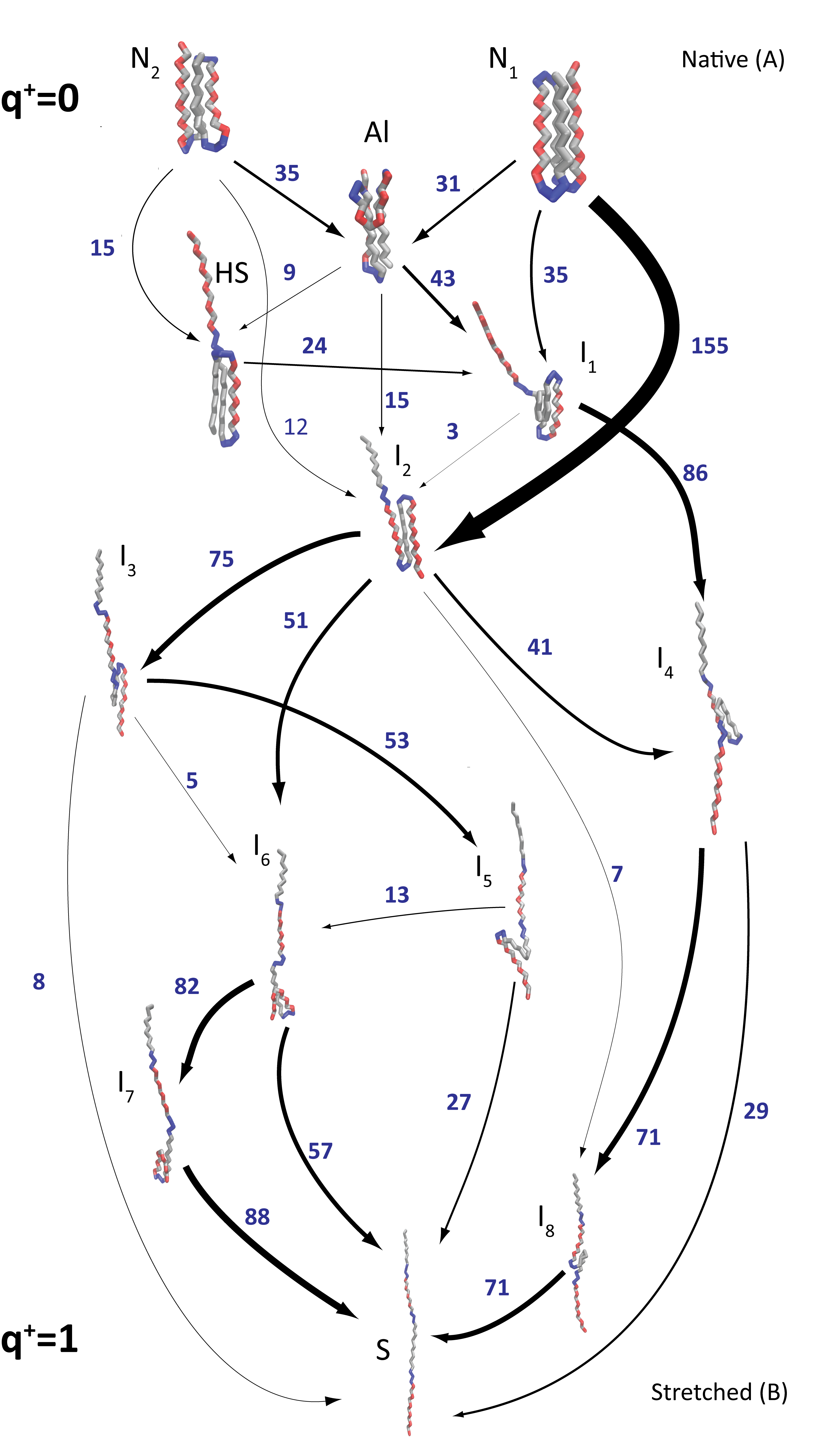}
\caption{\footnotesize Folding flux for the model protein. The network depicts the $85\%$ most relevant unfolding pathways for the $46-mer$ $BLN$ model protein
Each of the $13$ configuration identified with the Stochastic Steepest Descent algorithm are shown here, together with the label which identifies them. The configurations
are arranged vertically according to their committor probability (not in scale). The arrows connecting configurations represent the unfolding net flux $f^+_{ij}$, with their thickness is
proportional to the magnitude of the flux. The numbers next to the arrows give the flux magnitude in $10^{-9}ps^{-1}$.}
\label{flux}
\end{center}
\vspace{-0.6cm}
\end{figure}

We define the native set $A$ as basins $N_1$ and $N_2$, while the stretched set $B$ is just made up  of basin $S$. According to this definitions,
we calculate the committor probabilities, shown in Table \ref{tabla}. Figure \ref{flux} shows the net flux network, being the thickness of the arrows proportional
to the net flux $f^+_{ij}$. The total unfolding flux is $F=2.9\times 10^{-7}ps^{-1}$, meaning that we observe an unfolding transition every $3.5\mu s$, approximately.

\begin{figure}[h!]
\begin{center}
\includegraphics[width=0.5\textwidth]{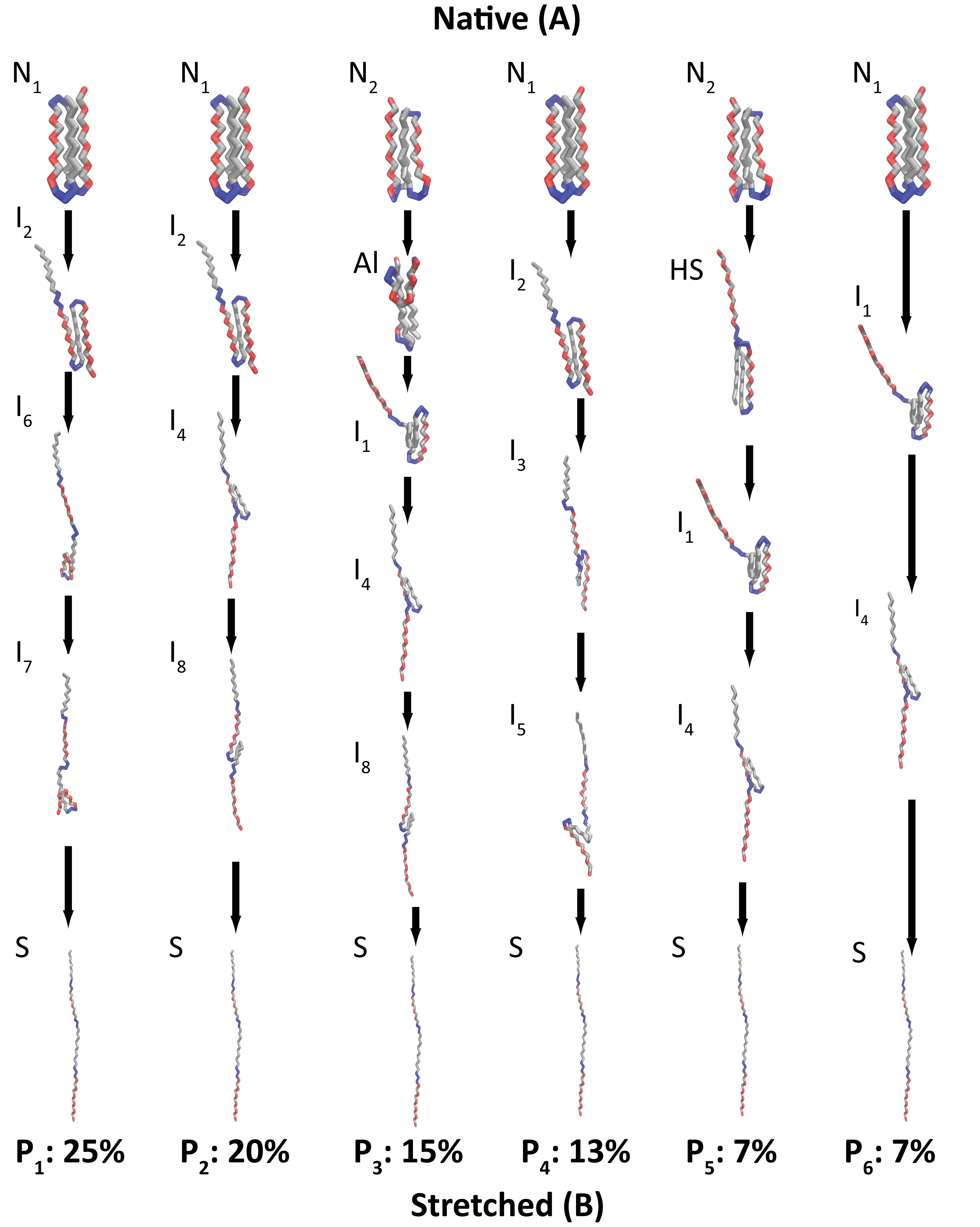}
\caption{\footnotesize Model protein unfolding pathways. The six pathways carrying most of the total flux (up to $89\%$) are explicitly shown.}
\label{paths}
\end{center}
\vspace{-0.5cm}
\end{figure}

We decompose the net flux network by identifying first the strongest pathway, remove it from the network and repeat the process until there is no path
from set $A$ to set $B$. Due to the size of our network, this process can be done manually, although computational applications can be used \cite{TPT3, TPT4}.
We identify a total of $9$ different paths leading from $A$ to $B$. After decomposing the network into these $9$ paths, unconnected regions still remain
due to the presence of \emph{trap states} \cite{TPT2} that carry around $20\%$ of the flux. Figure \ref{paths} shows the $6$ more relevant paths, 
which carry $89\%$ of the \emph{unfolding} flux.

From the $9$ pathways, $7$ start from conformation $N_1$ while just $2$ from $N_2$. This is a remarkable fact, being $N_1$ closer to the native structure
than $N_2$, as discussed in previous section. In addition, states $I_1$ and $I_2$ appear as the actual intermediates for the unfolding mechanism: $A\rightarrow B$ is forbidden in case
these two states are removed from the net flux network. Out of the $9$ pathways, $6$ of them pass through state $I_2$ and $3$ through state $I_1$.

The construction of the Markov Model from the PCs and the use of Transition Path Theory help us to unveil the actual unfolding mechanism and its driving process.
While HS is a notably relevant metastable state (indeed the most stable state under these conditions), its role in the unfolding mechanism is completely marginal, as
just appears in path $P_5$, with a weight of $7\%$. This important conclusion contradicts those derived from the one-dimensional description showed in  Fig. \ref{PMF}, where $HS$ was suggested as the  mechanical intermediate of the unfolding mechanism. The actual mechanical intermediates are $I_1$ and $I_2$ (not identified in the one-dimensional description), defining the two major unfolding routes. $I_2$ has a similar structure to $HS$, but while $HS$ maintains the hydrophobic core, in $HS$ the extremal $B_9$ strand is unfolded, breaking the core that stabilizes the structure and driving the unfolding mechanism. On the other hand, $I_1$ is more stable $\pi_{I_1}=0.07$ and represents a modified
$HS$ structure where the folded branches collapse into a globular structure which might lead to expose the extremal $B_9$ branch to the solvent and drive the unfolding mechanism
through states $I_4$ and $I_8$.

\section{Conclusions and Discussion}

In this paper we have presented the detailed analysis of the unfolding process of a model protein under the presence of a mechanical pulling force.
This scenario mimics force clamp single molecule experiments, where proteins or nucleic acids are subject to a constant external force that drives their unfolding.
Due to the limitation of available observables, these experiments are often analyzed by reconstructing their free-energy landscape along the pulling direction
through different existing techniques \cite{dudko06, Dudko2008, dudko11, review2014, app1, app2, app3}. This approach is often followed in many computational studies by using different reaction coordinates \cite{Q1, Q2, Best13, Best132, RMSD1, stock07}.

In this sense we wanted to reproduce a similar protocol and explore
the conclusions yielded by a one-dimensional analysis and a
multidimensional Markov model approach. The simplicity of our model
protein, and the fact that the force sets a privileged direction
invites to a one-dimensional characterization. Nonetheless, we have
seen how both approaches lead to contradictory conclusions. The PMF
description shows the existence of three major states, the native, the
stretched or denatured and a metastable Half-Stretched configuration
which seems to play the role of mechanical intermediate due to its
position in the free-energy profile.

Nonetheless, a more detailed multidimensional study changes
dramatically the unfolding picture. Being the most populate one, HS
state plays a marginal role in the unfolding pathway, with just $7\%$
of the unfolding flux passing through it. The true mechanical
intermediates are states $I_1$ and $I_2$, building the two major
unfolding routes, both related to the loss of the hydrophobic core
that destabilizes the structure and drives the unfolding process. In
this sense, due to the existence of multiple pathways, independently
of the chosen reaction coordinate, a one-dimensional picture would
\emph{never} be enough to characterize the unfolding pathway of this
system. Thus, our work differs from those which put attention on the
proper choice of the reaction coordinate \cite{multi2,
  review2014}. The necessity of multidimensional descriptions indeed
has been warned in the last years to understand thermal unfolding,
where the protein transits from a low-entropy state (native) to a
high-entropy one (denatured)~\cite{stock07, multi}. The
one-dimensional picture, however, is vastly assumed in mechanical
unfolding processes, both in experimental and computational
applications.

Regarding our analysis Markov Model protocol, we stress two major differences when compared to most works of this community. First, it is important to note
that we are actually using the PCs as reaction coordinates in order to reduce the system dimensionality. Nevertheless, these coordinates has been proven to capture
successfully the most relevant dynamical events of complex systems such as biomolecules. In our case, three coordinates are enough, as the remaining ones account merely
for gaussian thermal fluctuations. Second, we stress on the importance of the coarse-graining mechanism applied to the original Conformational Markov Network \cite{dpg},
which is able to systematically cluster the network based only on the kinetic properties of the system. 

Although extremely simple molecular assays such as DNA or RNA hairpins
could fit into a single reaction coordinate description~\cite{app1},
increasing slightly the complexity of the molecule leads to a
dramatical rise in the complexity of the actual free energy landscape
in the system, requiring more detailed studies. In this sense,
molecules such as multiple nucleic-acid hairpins~\cite{hairpin3},
protein-ligand complexes~\cite{dudkoclap} or any mechanically pulled
protein~\cite{jul}, appear as potential systems where a
one-dimensional description takes the risk of leading to a clear
misunderstanding of the actual complexity of their conformational
space and the dynamical processes to which they are subject.

\begin{acknowledgements}
The authors acknowledge support from the Spanish MINECO, project
FIS2011-25167 cofinanced by FEDER funds, and Gobierno de Arag\'on
(FENOL group).
\end{acknowledgements}

\begin{appendix}
\section{Model parameters and simulation protocol}
We simulate our system using the following adimensional parameters in Eq. (1),:
\begin{itemize}
	\item \textbf{$V_1$}: $K=50$, $r_0=1$.
	\item \textbf{$V_2$}: $A=5.118$, $B=5.308$, $V_0=-5.295$
	\item \textbf{$V_3$}: $C_i=0$ and $D_i=0.2$ if two or more aminoacids are neutral, and $C_i=D_i=1.2$ otherwise.
	\item \textbf{$V_4$}: there are three different cases, according to the character of the aminoacids.
		\begin{enumerate}
			\item $c_{ij}=0$ and $\epsilon_{ij}=4$ if $i$ or $j$ are neutral.
			\item $c_{ij}=1$ and $\epsilon_{ij}=4$ if $i$ and $j$ are hydrophobic.
			\item $c_{ij}=-1$ and $\epsilon=8/3$ in the remaining cases.
		\end{enumerate}
\end{itemize}
All simulations were carried out using self-built code, integrating the overdamped Langevin equations described above with an stochastic second order Runge-Kutta algorithm \cite{RK}. 

Physical units can be easily recovered in the following way. Length unit is defined by the $C_\alpha-C_\alpha$ distance $r_0=0.38nm$. Energy units are defined as the energy of a hydrogen bond $\epsilon_H\approx 1.7k_BT$, being force units $\tilde{F}\approx 17.3pN$. Mass unit is that of an average aminoacid $m_a\approx 3\times10^{-22}kg$. In this sense our time units $\tilde{t}=\sqrt{m_ar_0^2/\epsilon_H}\approx 3ps$, and the damping is that of water $\gamma\approx 10\frac{m_a}{\tilde{t}}$.

Six trajectories at $F=0.8F_U$ were simulated (with $F_U\approx 20pN$), were monomer $1$ was kept fixed while force was exerted to monomer $N$ through a linear spring.  Each simulation  covered a total time of $3ms$, with a previous thermalization process of $3\mu s$.  The integration step is $dt=0.005\tilde{t}$ and the time window to sample the trajectories $\tau=5\tilde{t}$.

\section{}

\subsection{Conformational Markov Network}
The Conformational Markov Network (CMN) \cite{cmn1, dpg} appears as a useful coarse-grained representation of large stochastic trajectories. This picture is obtained by discretizing the conformational space explored by the system and considering the dynamical jumps between the discretized configurations along the simulation. In this sense, the nodes of the complex network are defined by the discretized states, while the links account for the observed transitions between them. The arising network is thus a  weighted and directed graph. 

In our case, the conformational space is defined by the three first principal components, in order to reduce the number of degrees of freedom, keeping indeed the essential features of our system. We divide each of the principal component into $30$ cells of equal volume. Our discretized conformational space is thus made up of $30^3$ posible states, which may be or not occupied within the stochastic trajectory. We assign each node a weight $\pi_i$ accounting for the fraction of trajectory that the system has visited within the trajectory. The normalization condition $\sum_i\pi_i=1$ holds. Secondly, the value $T_{ij}$ is assigned to each directional link accounting for the dynamical jumps from node $j$ to $i$. Self-loops can exist, and thus $T_{ii}\neq 0$. Finally the normalization condition $\sum_iT_{ij}=1$ is forced. According to this, the CMN is totally defined by the occupancy vector $\Pi=P_i$ and the transition matrix $\tilde{T}=\{T_{ij}\}$. The matrix $\tilde{T}$ is the transition probability of the Markov chain defined by:

\begin{equation}
\Pi(t+\Delta t)=\tilde{T}\Pi(t),
\end{equation}
where $\Pi(t)$ it the probability distribution at time $t$. If the trajectory is long enough,  $\tilde{T}$ is ergodic and time invariant, vector $\Pi$ coincides with the stationary distribution associated with the Markov chain $\Pi=\tilde{T}\Pi$. Morover, the detailed balance condition must hold:

\begin{equation}
T_{ji}\pi_{i}=T_{ij}\pi_{j}.
\end{equation}

\subsection{Stochastic Steepest Descent}
Once we have translated de molecular dynamics trajectories onto a CMN, we apply the stochastic steepest descent (SSD) algorithm \cite{dpg} in order to split it into its basins of attraction in an efficient way, obtaining in turn useful thermo-statistical information about the system.The SSD algorithm is inspired in the deterministic steepest descent algorithm used to find minima in a multidimensional surface. We define the assisting vector $\textbf{U}=\{u_i\}$, where $i$ labels the nodes. The steps of the SSD algorithm are the following:

\begin{enumerate}
\item We start with $\textbf{U}=\textbf{0}$.
\item Select randomly a node $l$ with $u_l=0$ and write an auxiliary list of nodes adding $l$ as first entry.
\item Select within the neighbors of $l$ the node $m$ that follows the maximum probability flux, this is $T_{ml}=\max\{T_{jl, \forall j\neq l}\}$. Check which of the following conditions is fulfilled:
	\begin{enumerate}
		\item If $T_{ml}>T_{lm}$ and $u_m=0$, add $m$ to the list and go back to 3. using $m$ instead of $l$.
		\item If $T_{ml}>T_{lm}$ and $u_m\neq0$ write the labels of all the nodes in the list as $u_j=u_m$. Go back to step 3.
		\item If $T_{ml}\leq T_{lm}$ remove link $l \rightarrow m$ from the graph. Return to point 3. 
	\end{enumerate}
\end{enumerate}

This process ends when every node in the CMN has been labelled, this is $u_i\neq0 \, \forall \, i$. Then, the whole conformational space has been characterized and every node is connected with its local minima in the FEL. All nodes with the same label belong to the same basin in this FEL and therefore we can associate them with the same conformational state. 

Given the basin partition, a new CMN network can be built, taken the basins themselves as new nodes. The occupation probabilities will now be defined as $\pi_\alpha=\sum_{i\in \alpha} \pi_i$, while the  new transition matrix $\tilde{T}$ is built, with elements $T_{\beta\alpha}=\sum_{i\in\alpha}\sum_{j\in\beta}T_{ji}\pi_i/\sum_{i\in\alpha} \pi_i$. From these definitions, transition times can be easily calculated as $t_{\alpha\beta}=\tau/T_{\beta\alpha}$, being $\tau$ the time window used for the network construction. The relative free energy of basin $\alpha$ with respect to basin $\beta$ is simply $\Delta F_\alpha=-k_BT\log(\pi_\alpha/\pi_\beta)$.

\section{Thermal and mechanical characterization}

\begin{figure}[h!]
\begin{center}
\includegraphics[width=0.35\textwidth]{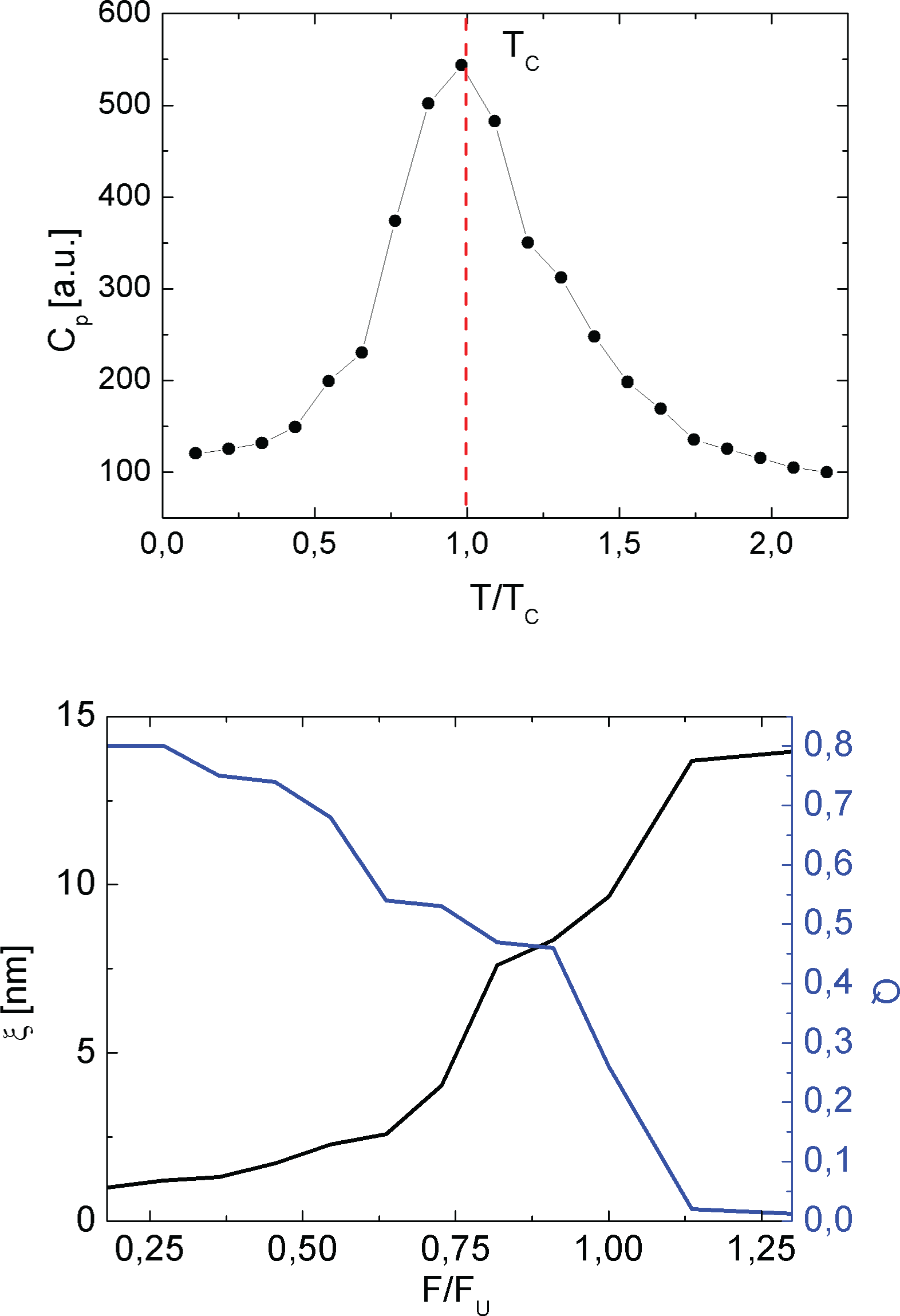}
\caption{\footnotesize Thermal and mechanical characterization of the model protein. At $T_c$ it exhibits a thermodynamical unfolding transition, reflected in a peak on the heat capacity (in arbitrary units). Force also induces an unfolding transition at $F_U$, leading to the fully stretched conformation.}
\label{char}
\end{center}
\end{figure}

We start by characterizing the protein from a thermal and mechanical point of view, in order to know the suitable range of force and temperature to work with. Although more detailed characterizations have been made in previous works \cite{model3} we focus on the thermodynamical transition at $T_c$, reflected on a peak in the heat capacity, as it can be seen in Fig. \ref{char}. The heat capacity is calculated as $C_p=(k_BT)^{-2}[\langle E^2\rangle -\langle E\rangle^2]$, with $E$ the total internal energy. We work at $T=0.55T_c$, below the transition, but with allowing enough fluctuation for the system to explore its configurational space.

When applying force to the protein, it exhibits also a transition at $F_U$, where the protein unfolds mechanically to the fully stretched configuration. At this force, the end-to-end distance $\xi$ increases abruptly, while the fraction of native contacts $Q$ drops to $0$. Around $F=0.75F_U$ a first change of behavior can be seen, due to the population of the half-stretched configuration, which leads to a drop to $Q\sim 0.5$ and $\xi\sim 7nm$.

\end{appendix}


\newpage
\begin{thebibliography}{55}


\bibitem{unf0} J. N. Onuchic and P. G. Wolynes, Curr. Opin.
  Struct. Biol. \textbf{14}, 70 (2004).

\bibitem{unf1} T. R. Sosnick and D. Barrick, Curr. Opin. Struct. Biol. \textbf{21}, 12 (2011). 

\bibitem{unf2} C. D. Snow, H. Nguyen, V. S. Pande, and M. Grubele, Nature \textbf{420}, 102 (2002). 

\bibitem{unf3} B. Onoa, S. Dumont, J. Liphardt, S. B. Smith, I. Tinoco Jr., and C. Bustamante, Science \textbf{299}, 1892 (2003). 

\bibitem{unf5} K. Lindorff-Larsen, S. Piana, R. O. Dror, and D. E. Shaw, Science \textbf{334}, 517 (2011).
		
		

\bibitem{OT1} J. Liphardt, B. Onoa, S. B. Smith, I. J. Tinoco and C. Bustamante, Science \textbf{292}, 733 (2001).

\bibitem{OT2} F. J. Ritort, J. Phys. C \textbf{18} R531 (2006).

\bibitem{MT1} W. J. Greenleaf, M. T. Woodside and S. M. Block, Annu. Rev. Biophys. Biomol. Struct. \textbf{36}, 171 (2007).
\bibitem{MT2} R. Liu, S. Garcia-Manyes, A. Sarkar, C. L. Badilla and J. M. Fern\'andez, Biophys. J. \textbf{96}, 3810 (2009).

\bibitem{AFM1} M. Carri\'on-V\'azquez, A. F. Oberhauser, S. B. Fowler, P. E. Marszalek, S. E. Broedel, J. Clarke, and J. M. Fern\'andez, Proc. Natl. Acad Sci. U.S.A. \textbf{97}, 3694 (1999).

\bibitem{AFM2} H. Li, A. F. Oberhauser, S. B. Fowler, J. Clarke and J. M. Fern\'andez, Proc. Natl. Acad. Sci. U.S.A. \textbf{97}, 6527 (2000).

\bibitem{AFM3} R. B. Best, S. B. Best, J. L. Toca-Herrera and J. Clarke, Proc. Natl. Acad. Sci. U.S.A. \textbf{99}, 12143 (2002).
	


\bibitem{FRET1} B. Schuler, E. A. Lipman and W. A. Eaton, Nature \textbf{419}, 743 (2002).
\bibitem{FRET2} B. Schuler and W. A. Eaton, Curr. Opin. Struct. Biol. \textbf{18} 16 (2008).  
\bibitem{dudko06} O. K. Dudko, G. Hummer and A. Szabo, Phys. Rev. Lett. \textbf{96}, 108101 (2006).
\bibitem{Dudko2008} O. K. Dudko, G. Hummer, and A. Szabo, Proc. Natl. Acad. Sci. U. S. A. {\bf 105}, 15755 (2008).
\bibitem{dudko11} O. K. Dudko, T. G. W. Graham, and R. B. Best, Phys. Rev. Lett. \textbf{107} 208301, (2011)
\bibitem{review2014} M. T. Woodside, and S. M. Block, Annu. Rev. Biophys. \textbf{43} 19, (2014)


\bibitem{Q1} P. G. Wolynes, Q. Rev. Biophys. \textbf{38}, 405 (2005).
\bibitem{Q2} P. G. Wolynes, J. N. Onuchi and D. Thirumalai, Science \textbf{267}, 1619 (1995).
\bibitem{Best13} R. B. Best, G. Hummer and W. A. Eaton, Proc. Natl. Acad. Sci. U.S.A  \textbf{110}, 17874 (2013)
\bibitem{Best132} E. R. Henry, R. B. Best and W. A. Eaton, Proc. Natl. Acad. Sci. U.S.A. \textbf{110}, 17880 (2013).

\bibitem{RMSD1} S. Piana, K. Lindorff-Larsen and D. E. Shaw, Proc. Natl. Acad. Sci. U.S.A. \textbf{110} 5915 (2012).
\bibitem{stock07} A. Altis, P. H. Nguyen, R. Hegger and G. Stock, J. Chem. Phys. \textbf{126}, 244111 (2007).
\bibitem{PCA92} A. E. Garcia, Phys. Rev. Lett. \textbf{68}, 2696 (1992).
\bibitem{PCA93} A. Amadei, A. B. M. Linssen and H. J. C. Berendsen, Proteins \textbf{17}, 412 (1993).
\bibitem{pca2d} G. G. Maisuradze, A.  Liwo, and H. A. Scheraga, Phys. Rev. Lett. \textbf{102}, 238102 (2009)


\bibitem{BLN1} J. D. Honeycutt, D. Thirumalai, Proc. Natl. Acad. Sci. U.S.A. \textbf{87}, 3526 (1990)
\bibitem{BLN2} J. D. Honeycutt, D. Thirumalai, Biopolymers \textbf{32} 695 (1992)
\bibitem{modelor} S. Brown, N. J. Fawzi, T. Head-Gordon, Proc. Natl. Acad. Sci. U.S.A. \textbf{100}, 10712 (2003).
\bibitem{modelor2} S. Brown and T. Head-Gordon, Protein Sci. \textbf{13}, 958 (2004).
\bibitem{model} D. J. Lacks, Biophys. J. \textbf{88}, 3494  (2005).
\bibitem{wales3} D. J. Wales and P. E. J. Dewsbury, J. Chem. Phys. \textbf{121}, 10284 (2004)
 \bibitem{wales4} M. A. Miller, D. J. Wales, J. Chem. Phys.  \textbf{111}, 6610 (1999)
\bibitem{model2} A. Imparato, S. Luccioli and A. Torcini. Phys. Rev. Lett. \textbf{99}, 168101 (2007).
\bibitem{model3} S. Luccioli, A. Imparato, S. Mitternacht, A. Irb\"ack and A. Torcini, Phys. Rev. E \textbf{81}, 010902(R) (2010).
\bibitem{wales2} D. J. Wales and T. Head-Gordon,  J. Phys. Chem. B \textbf{116}, 8394-8411 (2012)









\bibitem{noebook} G. R. Bowman, V. S. Pande and F. No\'e (Eds.)  \textit{An Introduction to Markov State Models and Their Application to Long Timescale Molecular Simulation}. Advances in Experimental Medicine and Biology. (2014).	
\bibitem{pande1} S. J. Klippenstein, V. S. Pande and D. G. Truhlar, J. Amer. Chem. Soc., \textbf{136}, 528 (2014).

	
\bibitem{cmn1} F. Rao, and  A. Catfisch, J. Mol. Biol. \textbf{342}, 299(2004).
\bibitem{dpg} D. Prada-Gracia, J. G\'omez-Garde\~nes, P. Echenique, and F. Falo, PLoS Comput. Biol. \textbf{5}, e1000415 (2009).




\bibitem{TPT1} E. W. Vanden-Eijnden J. Stat. Phys. \textbf{123} 503 (2006)
\bibitem{TPT2} F. Noe, C. Schutte, E. Vanden-Eijnden, L. Reich and T.R. Weikl, Procc. Netl. Acad. Sci. U.S.A, \textbf{106}, 19011-19016 (2009)
\bibitem{TPT3} P. Metzner, C. Shutte, E. Vanden-Eijnden, Multiscale Model. Simul. \textbf{7} 1192 (2009)
\bibitem{TPT4} R. Banerjee and R. I. Cukier, J. Phys. Chem. B, \textbf{118}, 2883 (2014)




	
	


\bibitem{multi} S. V. Krivov and M. Karplus, Proc. Natl. Acad. Sci. U.S.A. \textbf{101}, 14766 (2004)
\bibitem{multi2} O. K. Dudko, T.G. W. Graham and R. B. Best, Phys. Rev. Lett. \textbf{107}, 208301 (2011)


\bibitem{app1} J. Liphardt, B. Onoa, S. B. Smith, I. Tinoco Jr, C. Bustamante, Science \textbf{292} 733 (2001). 
\bibitem{app2} G. Hummer and A. Szabo, Proc. Natl. Acad. U.S.A. \textbf{98} 3658 (2000)
\bibitem{app3} M. Li, A M. Gavovich and A. I. Voitenko, J. Chem. Phys. \textbf{129}, 105102 (2008)



\bibitem{gephi} M. Bastian, S. Heymann and M. Jacomy. International AAAI Conference on Weblogs and Social Media. (2009).
\bibitem{modul} V. D. Blondel, J. L. Guillaume, R. Lambiotte and E. Lefebvre, J. Stat. Mech: Theor. and Exp. \textbf{10}, P1000 (2008).




\bibitem{hairpin3} A. Alemany, A. Mossa, I. Junier and F. Ritort, Nature Physics \textbf{8}, 688 (2002).
\bibitem{dudkoclap} Y. Suzuki and O.K. Dudko, Phys. Rev. Lett. \textbf{110}, 158105 (2013).
\bibitem{jul} J. Alegre-Cebollada, \emph{et. al.} Cell \textbf{156}, 1235  (2014).
\bibitem{RK} H. S. Greenside and E. Helfand, Bell Syst. Tech. J. \textbf{60}, 1927 (1981)
	

\end{thebibliography}
\end{document}